\documentclass[conference,a4paper]{IEEEtran}
\usepackage{graphicx}
\usepackage{epstopdf}
\usepackage{cite}
\usepackage{amsfonts}
\usepackage{amssymb}
\usepackage{amsmath}
\usepackage{mathrsfs}
\usepackage{bm}
\usepackage{amsmath}
\usepackage{mdwmath}
\usepackage{mdwtab}
\usepackage[ruled,vlined,linesnumbered]{algorithm2e}

\begin{document}

\title{Remote Channel Inference for Beamforming in Ultra-Dense Hyper-Cellular Network}
\author{\IEEEauthorblockN{Sheng Chen$^1$, Zhiyuan Jiang$^1$, Jingchu Liu$^1$, Rath Vannithamby$^2$, Sheng Zhou$^1$,  \\Zhisheng Niu$^1$, \emph{Fellow, IEEE} and Ye Wu$^3$}
    \IEEEauthorblockA{$^1$Tsinghua National Laboratory for Information Science and Technology,\\
        Department of Electronic Engineering, Tsinghua University, Beijing, 100084, China\\
        $^2$ Intel Corporation, Hillsboro, Oregon,\\
        $^3$ ICRI-MNC, Intel Labs, Intel Corporation,\\
        \{chen-s16, liu-jc12\}@mails.tsinghua.edu.cn, \{zhiyuan, sheng.zhou, niuzhs\}@tsinghua.edu.cn,\\
        \{rath.vannithamby, ye.wu\}@intel.com}}
\maketitle

\begin{abstract}
In this paper, we propose a learning-based low-overhead channel estimation method for coordinated beamforming in ultra-dense networks. We first show through simulation that the channel state information (CSI) of geographically separated base stations (BSs) exhibits strong non-linear correlations in terms of mutual information. This finding enables us to adopt a novel learning-based approach to remotely infer the quality of different beamforming patterns at a dense-layer BS based on the CSI of an umbrella control-layer BS. The proposed scheme can reduce channel acquisition overhead by replacing pilot-aided channel estimation with the online inference from an artificial neural network, which is fitted offline. Moreover, we propose to exploit joint learning of multiple CBSs and involve more candidate beam patterns to obtain better performance. Simulation results based on stochastic ray-tracing channel models show that the proposed scheme can reach an accuracy of $99.74\%$ in settings with $20$ beamforming patterns.

\end{abstract}

\section{Introduction}
\begin{figure}[!t]
    \centering
    \includegraphics[width=0.45\textwidth]{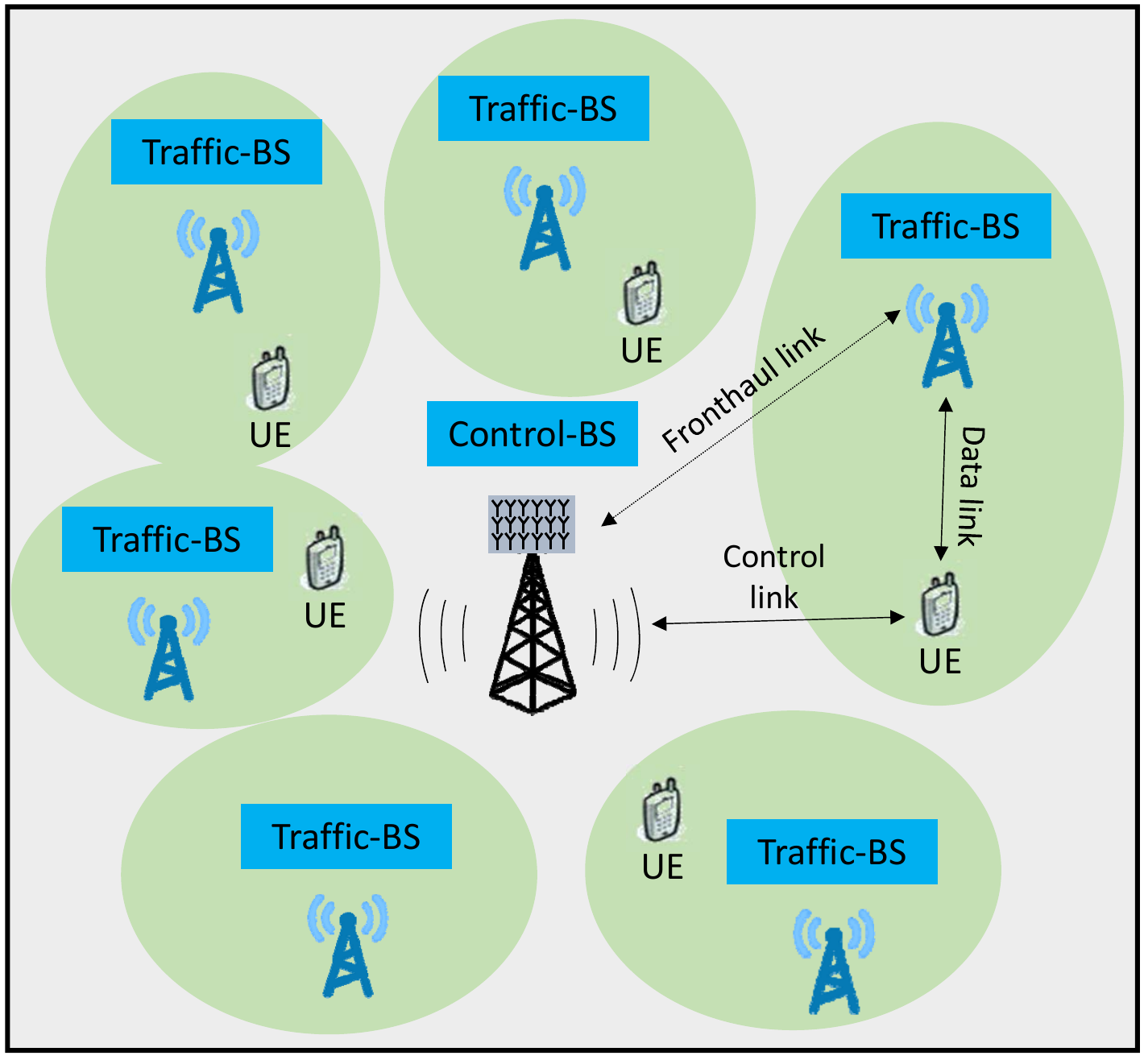}
    \caption{An ultra-dense HCN, with one control-BS and several traffic-BSs.}
    \label{Fig_sys_arch}
\end{figure}

In order to meet the rapidly growing demand for wireless communication services, several novel network architectures have been proposed to increase system capacity and improve the quality of service. Among them, network densification and software-defined virtualization are considered as key technologies for future systems \cite{Andrews14}. The basic idea behind the \emph{ultra-dense network} concept is to reduce the transmission distance between users and base stations (BSs), which yields to less pathloss, more spectrum options and hence higher data rate. Meanwhile, the \emph{hyper-cellular network} (HCN) is proposed in \cite{niu12} \cite{zhou2016software}, based on which the control- and data-channel coverages are separated to provide uniform high quality of service with elastic data-channel coverage. Fig. \ref{Fig_sys_arch} shows a sketch of an ultra-dense HCN, where one control-BS (CBS) is responsible for control coverage and several traffic-BSs (TBSs) take care of data traffic. This architecture allows agile sleeping of TBSs, and thus the energy consumption of the network can be drastically reduced, while the CBS can handle the control signaling to provide seamless coverage.

A major technical challenge for ultra-dense HCN is the strong interference among densely deployed TBSs \cite{hwang13}. Coordinated beamforming is an effective way to mitigate interference \cite{bartoli14}. However, it may encounter a big hurdle for timely and accurate channel state information (CSI) acquisition. Generally, the system exploits dedicated pilot signals to acquire the CSI. However, on one hand, limited pilot resources cannot satisfy the demand of a huge number of densely deployed TBSs. On the other hand, searching over all possible pilot sequences of TBSs leads to increased terminal energy consumption and time overhead. Therefore, novel efficient CSI acquisition methods are needed\footnote{Note that in this paper, we consider the frequency-division duplex (FDD) systems without channel reciprocity, and therefore the pilot length must scale with the number of BSs. Although time-division duplex is also a viable solution, the FDD systems cannot be overlooked due to dedicated spectrum and wide usage.}.

Extensive work has been devoted to reducing pilot training overhead for CSI acquisitions. The work in \cite{Adhikary13,Jiang14,jiang17,Choi142}, proposes to leverage spatial channel correlations among co-located BS antennas to reduce the overhead and shows considerable performance gain. A large body of work adopts the hybrid (analog and digital) beamforming methods, in which the analog beams are shaped with long-term channel statistics \cite{molisch04mag}. Considering analog beam generation, beam sweeping is one of the traditional beamforming schemes that save pilot resources at a cost of long training time \cite{wong01}. Some other work uses multi-level search or heuristic algorithms to reduce training time \cite{tsang11} \cite{li13}. Even so, beam sweeping algorithms still need non-negligible training time and overhead. Authors in \cite{kela16} propose a location-based beamforming scheme for UDN, by which cells estimate the locations of user equipments (UEs) based on uplink beacons and generate beamforming vectors according to location information. However, this scheme does not perform well when the line-of-sight (LoS) path does not exist. Furthermore, privacy problems may also occur in location-aware schemes. To the best of our knowledge, the existing literature seldom explores the possibility that the CSI of one BS can be inferred by the CSI of neighbouring BSs, because it is generally believed that the CSI of geographically separated BSs is independent of each other. Our previous work \cite{liu15} challenges this traditional belief with the \emph{channel learning} framework, in which supervised learning is exploited to make cross-BS channel inferences for the purpose of BS selection. Obviously beamforming needs more accurate CSI compared with BS selection, and hence new approaches are needed.

In this paper, we propose a remote beamforming inference framework to address the coordinated beamforming problem in ultra-dense HCN. The main contribution is that the existence of \emph{nonlinear} correlations between the CSI of geographically separated BSs is shown by calculating the mutual information of these variables. This finding motivates a learning-based approach for mining the non-linear relationship for remote channel inference. Specifically, we first adopt the supervised learning approach to train an artificial neural network (NN) offline with CBS-TBS CSI data pairs obtained from pilot-aided channel training, so that the NN learns the non-linear mapping between remote channels. Then we employ the learned NN to select beam patterns for the TBS solely based on CBS CSI, which is estimated online with pilots. Simulation results show that our scheme has good prediction accuracy and adapts well to different environments based on a widely-used geometrical stochastic channel model (GSCM) \cite{molisch04_gscm}.

\section{System Architecture}
\label{section2}
We consider a HCN with a CBS and several TBSs. The CBS is equipped with $N_\textrm{m}$ antennas while each TBS has $N_\textrm{s}$ antennas.

\subsection{Problem Formulation}
In our proposed remote channel inference framework under HCN, in order to reduce pilot overhead, only the CBS sends downlink reference signals. Therefore, only the CBS can be aware when user-equipments (UEs) start searching available network for attachment. After a UE is attached to the CBS, the central controller at CBS will decide which TBS is going to serve the UE. Then the selected TBS will exploit beamforming technology to transmit downlink data. However, the CSI knowledge is necessary for beamforming but obtaining the exact CSI between the TBS and the UE is believed to be extremely difficult without the TBS transmitting downlink pilots in an FDD system. We resort to the data-driven channel learning based approach and use the CSI between the CBS and UEs to infer the channels between TBSs and UEs.

According to the above analysis, our goal is to estimate the CSI of TBSs for evaluating the optimal transmit beamforming vector $\bm{w_\textrm{s}}$ given the CSI $\bm{h_\textrm{m}}$ at remote CBS. Here $\bm{h_\textrm{m}}$ is a $N_\textrm{m} \times 1$ vector and $\bm{w_\textrm{s}}$ is a normalized $N_\textrm{s} \times 1$ vector, which are both complex-valued.

\subsection{Channel Model}
We use GSCM \cite{molisch04_gscm} as our channel model. In this model, channel of the direct path between one UE and one antenna can be written as:
\begin{equation}
  h_{\textrm{los}}=\frac{\lambda }{2\pi d_{0}} \beta_0 \exp{\left(-\frac{j2\pi d_{0}}{\lambda}\right)},
\end{equation}
where $\lambda$ is the carrier wavelength, $\beta_0$ is the amplitude, and $d_0$ is the length of the direct path. As for the non-LOS paths, channel response can be written as:
\begin{equation}
  h_{\textrm{nlos}}=\sum_{i}^{N_{\textrm{sca}}} \frac{\lambda }{2\pi d_{i}} \beta_i \exp{\left(-\frac{j2\pi d_{i}}{\lambda}+j{\phi}_i\right) },
\end{equation}
where $d_{i}$ is the length of the $i$-th non-LOS path, $\beta_i$ is the amplitude of path $i$, and ${\phi}_i$ represents the random phase shift due to scattering. Given a Rician factor of $K$, summing these components up obtains the channel response:
\begin{equation}
  h=\sqrt{\frac{K}{K+1}}h_{\textrm{los}}+\sqrt{\frac{1}{K+1}}h_{\textrm{nlos}}.
\end{equation}
The channel response at each antenna is calculated to form a channel vector $\boldsymbol{h}$. Given the normalized beamforming vector $\boldsymbol{w}$ and transmit signal $x$, receive signal of a single link can be written as:
\begin{equation}
  y = {\boldsymbol{h}}^{H}\boldsymbol{w}x+n,
\end{equation}
where $n$ is the additive Gaussian noise. Ignoring the interference from other UEs\footnote{Multiple UEs will be considered in the simulations}, optimal performance can be achieved by maximum ratio transmission (MRT) scheme, i.e., $\boldsymbol{w}=\boldsymbol{h}$.

\section{Remote Beamforming Inference}
\label{section3}
The optimal MRT beamforming structure, or eigen-based beamforming scheme where only channel statistics are available \cite{lo99}, requires accurate channel estimations to derive the beamforming vector (pattern). In the proposed framework, we use remote channel inference to find the optimal beamforming vector for TBS. To simplify the approach, a finite, prescribed codebook is involved, and the task is to select the beamforming vector (codeword) out of the codebook. With the above simplification, the problem is transformed into finding the optimal beam pattern for TBS from the prescribed codebook according to the observable CSI of CBS. Here we define the beamforming pattern that has the maximum signal to noise ratio (SNR) as the optimal one. Before giving more details about our beamforming inference scheme, the following subsection is dedicated to show that non-linear correlations exist between the CSI of CBS and the optimal beamforming vector of TBSs, which justifies the channel inference feasibility in our proposed framework. Most existing literature assumes the CSIs from geographically separated BSs are statistically independent by showing that their linear correlation is approximately zero. However in this work, we show that they are actually highly dependent with high \emph{non-linear} correlation in terms of mutual information based on the GSCM model.

\subsection{Non-Linear Correlation Analysis for Remote Channels}
To prove the existence of this correlation, we first give some high-level intuition. The optimal beamforming vector at TBS is deterministic when the CSI at TBS is given. Moreover, CSI is a function of UE locations. This function changes with time, but it can be assumed that the CSI is quasi-static as long as the estimation time is smaller than channel coherence time. Based on these two facts, it can be inferred that the optimal beamforming vector should be strongly related to UE locations, especially when the LoS path exists. Similarly, UE locations are also strongly related to the observable remote CSI of the CBS. Under the condition that CBS antennas form a large scale antenna array, existing literature shows that the channel responses from different UE locations are asymptotically orthogonal \cite{Marzetta10}. Thus the mapping from UE location to CBS array response is invertible. Therefore, the optimal beamforming pattern at TBSs can be inferred by first estimating the UE location from CBS channel and then use the estimated location to find the beam pattern at TBS. Moreover, these two steps can be combined by learning the implicit mapping between CBS channel and TBS channel. In general, this implicit relationship is non-linear, and thus it does not exhibit any linear correlation.

\begin{figure}[!t]
\centering
\includegraphics[width=3.5in]{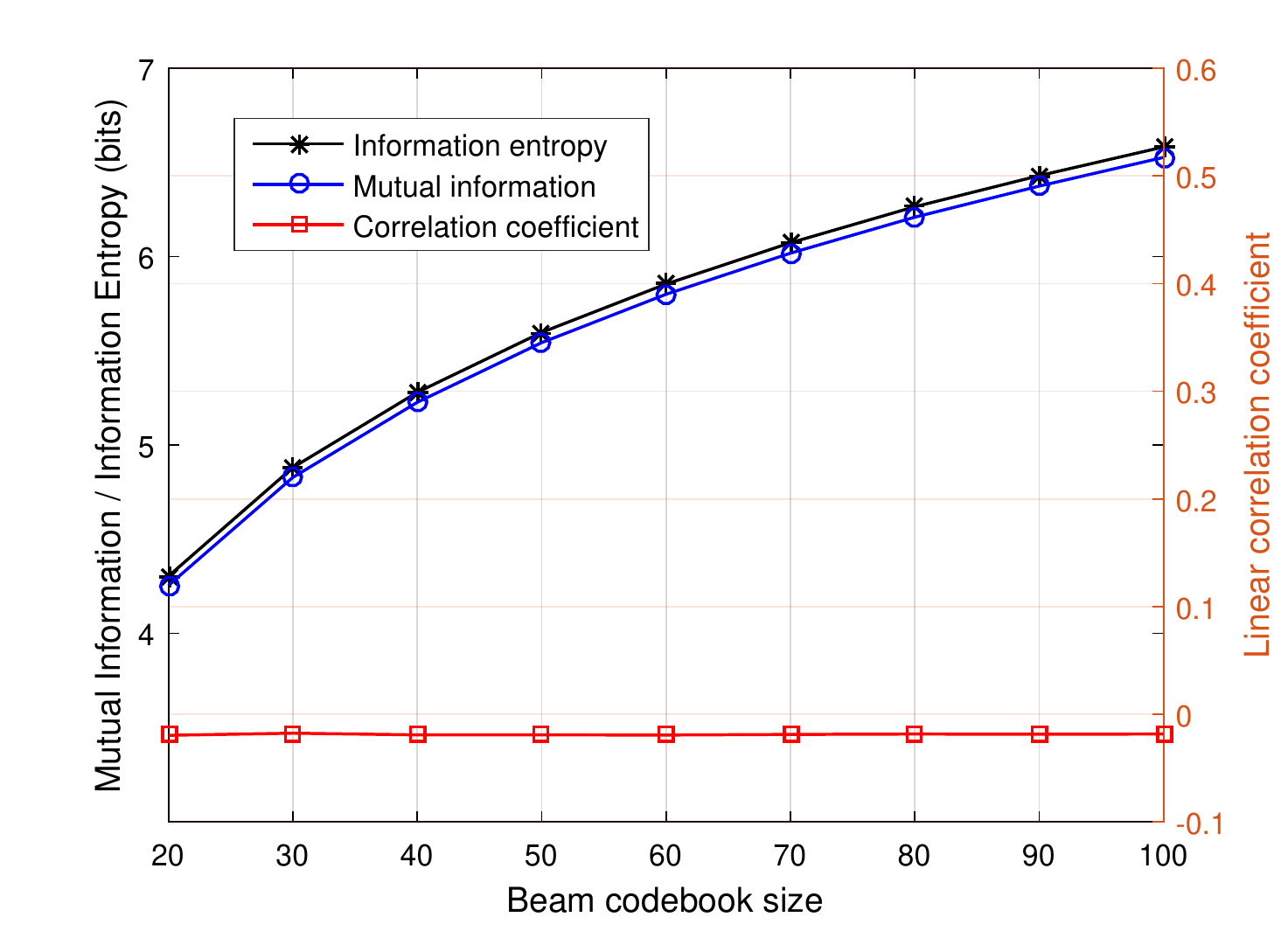}
\caption{Analysis on the correlations between observable CSI and remote beamforming pattern. The mutual information between observable CSI and remote beamforming pattern is close to the information entropy of beamforming pattern, while the linear correlation coefficient of these two variables is approximately zero.}
\label{sim_info}
\end{figure}

However, this relationship can be shown clearly by calculating mutual information, which measures the degree of relevance between two variables, between the observable CSI of CBS and remote optimal beam pattern. The calculation is based on a data-driven approach, whereby channel data are generated according to the channel model introduced before. Preprocessing such as principal component analysis (PCA) and quantization are employed to reduce dimensionality, and this preprocessing results in a conservative approximation of the real value according to \emph{the information processing theorem}\cite{Cover12}. Fig. \ref{sim_info} shows the estimated mutual information under different codebook size. Sinusoidal beamforming patterns are used here as codewords to obtain a variable angular resolution. It can be seen that the mutual information between the quantized amplitude of observable CSI in angular domain and the selected beam pattern is very close to the information entropy of the latter one, which means that the optimal beam pattern is almost certain given the remote CSI of CBS.

\begin{figure*}[!t]
    \centering
    \includegraphics[width=0.9\textwidth]{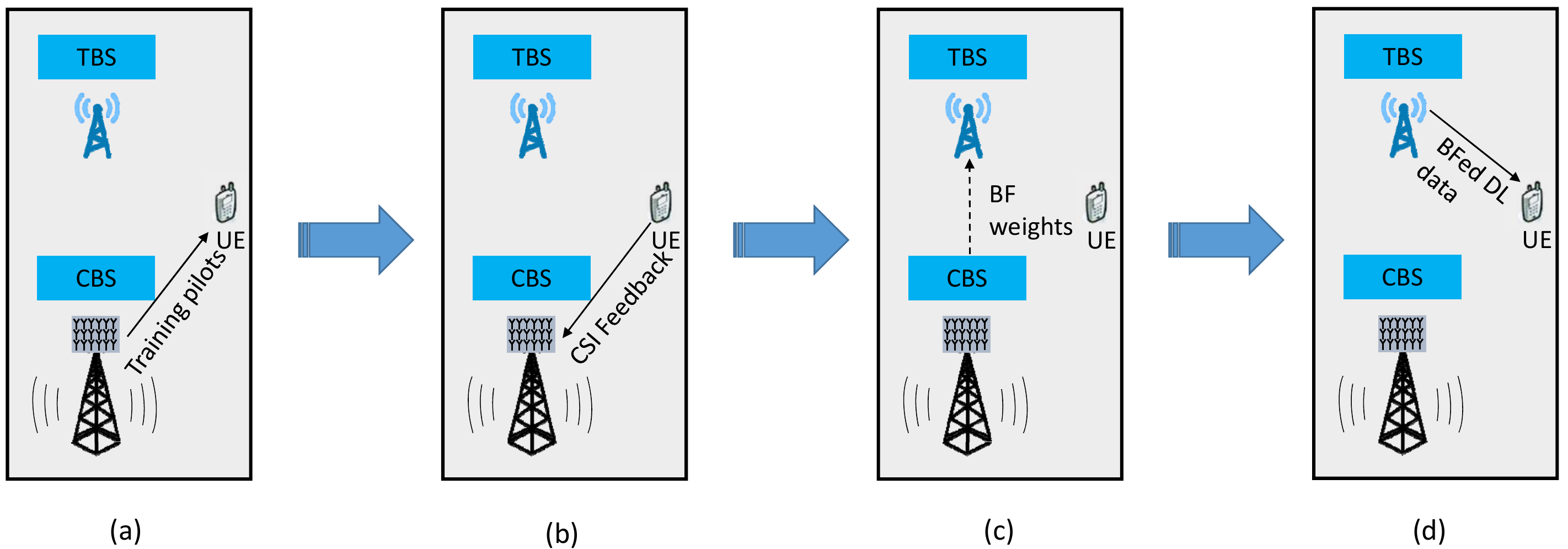}
    \caption{The high-level procedure of the proposed remote beamforming (BF) inference scheme. The CBS first broadcasts channel training pilots. Then the UE estimates downlink CSI and sends CSI feedback to the CBS. After that, the CBS generates beamforming weights for the TBS using a learning-based approach and transmits them to the TBS. Finally the TBS uses the beamforming weights to establish the link to the UE.}
    \label{RBFI}
\end{figure*}

\subsection{Procedures of Remote Beamforming Inference}
In this part, we introduce our end-to-end beamforming inference algorithm. Here end-to-end scheme means some intermediate variables such as UE locations will not be introduced.

As shown in Fig. \ref{RBFI}, our proposed scheme has four main procedures. The CBS first broadcasts channel training pilots. When a UE starts searching the available network for attachment, only the CBS can be aware. Consequently, the UE connects to the CBS and sends estimated downlink CSI feedback to the CBS. After that, the CBS selects beam pattern for the TBS using a learning-based approach. Finally, the TBS creates an initial connection to the UE using selected beam pattern. After the link to UE is established, TBS can send pilots through a dedicated beam for beam tracking or link adaption. If the connection fails in the last stage, TBS can ask for an updated beam pattern from the CBS or use a high-overhead default beamforming mechanism to finally establish the link. The above situation occurs when CSI changes during the prediction time or the predicted beamforming pattern deviates too much from the optimal one.

\subsection{Neural-Network-based Learning Approach}
It is quite difficult to obtain the exact form of the mapping between CSI at CBS and beamforming vector at TBS due to the complex structure of the physical environment. Therefore, we consider using neural networks, which are shown to achieve tremendous success in approximating non-linear functions \cite{hornik89}. The training of the neural network is done as follows. For a certain cellular network, channel estimations for both the CBS and TBSs can be done at sampling points, e.g., by pilot-aided channel training. After that training samples for the neural network can be generated from these sampling points.

The input of the neural network is a pre-processed version of the observable CSI at CBS while the output are the selection probabilities of beamforming patterns for TBSs. Since the CSI of TBSs is known for each training sample, it is trivial to find the optimal beam pattern from the codebook based on the training data. Therefore, the number of labels for this classifier is the same as the codebook size. When it comes to input, preprocessing is needed to extract features. Specifically, we transform channel vectors into angular domain by fast Fourier transforms, extract amplitude information, take the logarithm and use the Lloyds algorithm \cite{lloyd82} for each angular dimension to obtain quantized features. The size of input features is the same as the number of CBS antennas. As for the cost function, the first part is the cross-entropy \cite{golik2013cross}, which can be formalized as
\begin{equation}\label{costfunction}
  L = -\frac{1}{N}\sum_{i=1}^{N}\sum_{j=1}^{K}(y_{ij}\log(r_{ij})+(1-y_{ij})\log(1-r_{ij})).
\end{equation}
Here $N$ is the number of samples and $K$ is the number of labels. $y_{ij}$ and $r_{ij}$ represent the true value and predicted value of the $j$-th label for the $i$-th sample, respectively. The second part is the regularization term, which can be defined as
\begin{equation}\label{regularization}
  \lambda\sum_{m=1}^M\omega_m^2,
\end{equation}
where $\lambda$ is the regularization parameter and $\omega_m$\ is the $m$-th weight parameter of the neural work. We use back-propagation algorithm to calculate gradients and obtain the weight update based on those gradients.

After the neural network has been trained, predictions can be made to select optimal beam patterns. We feed extracted features into the network and the output vector stands for the selection probability of all the beam patterns. The beamforming vector with the largest probability is the output decision. The neural network is trained offline, while the low-complexity beamforming inference is made online, and thus the proposed scheme can reduce time overhead in comparison with other schemes like beam sweeping. 

\subsection{Enhancements for Prediction Accuracy}
\label{section4}
We have mentioned that the system performance relies on the prediction time and prediction accuracy. Since the offline training time is not a bottleneck due to the previous analysis, accuracy becomes the key factor. In this part, some practical methods are introduced to improve the prediction accuracy.

\subsubsection{Joint learning of multiple CBSs}
UE locations have a big influence on the system performance. Specifically, UEs that have shorter distances to TBSs tend to have worse performance. The reason is that the estimation error of angle of departure is inversely proportional to the distances between UEs and TBSs given certain estimation error of UE locations. In extreme cases that UEs are very close to TBSs, beam pattern selection can be a really difficult decision based on our framework. For these UEs, wider beam can be a good choice to trade angular resolution for the robustness of estimation..

According to the previous analysis, providing more information about UE locations can help improve prediction accuracy. To achieve this, one feasible solution is to include more CBSs into the system. In long-term-evolution (LTE) standards, UEs can measure neighboring BSs and send CSI reports to the attached one. In that way, the CBS can have CSIs of UEs to other CBSs and thus it can exploit this information to improve beamforming accuracy and hence the system performance.

\subsubsection{More Candidate Beams}
 A set of candidate beam patterns can be obtained through the learning approach. Specifically, the output of the neural network is soft information indicating the probability for selecting all beam patterns. If however the predicted beam pattern fails to work, the beam with the second highest selection probability can be a choice. Moreover, to cope with problems that the prescribed codebook may be unable to provide sufficient angular resolution, we can combine our work with existing beam sweeping scheme \cite{wong01}. We first use beamforming inference scheme to reduce the size of candidate list and then use training methods to find the optimal beam pattern. Either way, our proposed scheme can reduce, if not eliminate completely, the training overhead and training time.


\section{Simulations}
\label{section5}
In this section, numerical simulations are conducted to evaluate the performance of the proposed remote beamforming inference scheme. First the simulation settings are introduced and then simulation results are shown. We also compare our proposed scheme with location-based beamforming scheme and analyze the influences of the codebook size and number of scatterers.

\subsection{Simulation Settings}
A CBS with uniform linear array and half wavelength spacing is considered. Without lose of generality, a TBS with uniform linear array and half wavelength spacing is considered\footnote{Interference from other TBSs will be considered in future work.}. $20000$ UEs are randomly placed under the coverage of the TBS following uniform distribution. We use the channel model introduced in section \ref{section2} to generate CSI. Specifically, we randomly place several scatterers and calculate channel responses of direct paths and reflection paths from UEs to the CBS and TBSs. Then the weighted sum of line-of-sight component and non-line-of-sight components by a given Rician factor is obtained.  Values of other simulation parameters are listed in Table \ref{sim_parm}.

\begin{table}[!t]
\renewcommand{\arraystretch}{1.3}
\caption{Simulation Parameters}
\label{sim_parm}
\centering
\begin{tabular}{|c|c|}
\hline
\textbf{Parameter}& Value\\
\hline
\hline
Carrier frequency & 3.5 GHz\\
\hline
CBS antenna number & 100\\
\hline
TBS antenna number & 20\\
\hline
Array spacing & 0.0428 m\\
\hline
CBS coverage radius & 700 m\\
\hline
TBS coverage radius & 200 m\\
\hline
Rician factor & 10 dB \\
\hline
\end{tabular}
\end{table}

A neural network with one hidden layer is adopted to accomplish the training and prediction task. The size of the hidden layer is the same as the number of CBS antennas. The non-linearity function for the hidden layer is \emph{sigmoid}, while the non-linearity function for the output layer is \emph{softmax}. The input features are quantized CSI in the angular domain while the output are the selection probabilities of all beam patterns from a codebook. Here the prescribed codebook is the discrete-Fourier-transform (DFT) codebook, and hence the size of the codebook is the same as the number of TBS antennas. We randomly partition data set into three parts for training (70\%), validation (15\%) and test (15\%). For each sample, the performance metric is the received signal strength using selected beam pattern divided by that using optimal beam pattern from the codebook. We use this as our performance metric along with prediction accuracy because there is a chance that UEs have almost the same received signal strength using two beamforming vectors, in which case the optimal beamforming pattern can be either one of them.

\subsection{Simulation Results}
\begin{figure}[!t]
\centering
\includegraphics[width=3.5in]{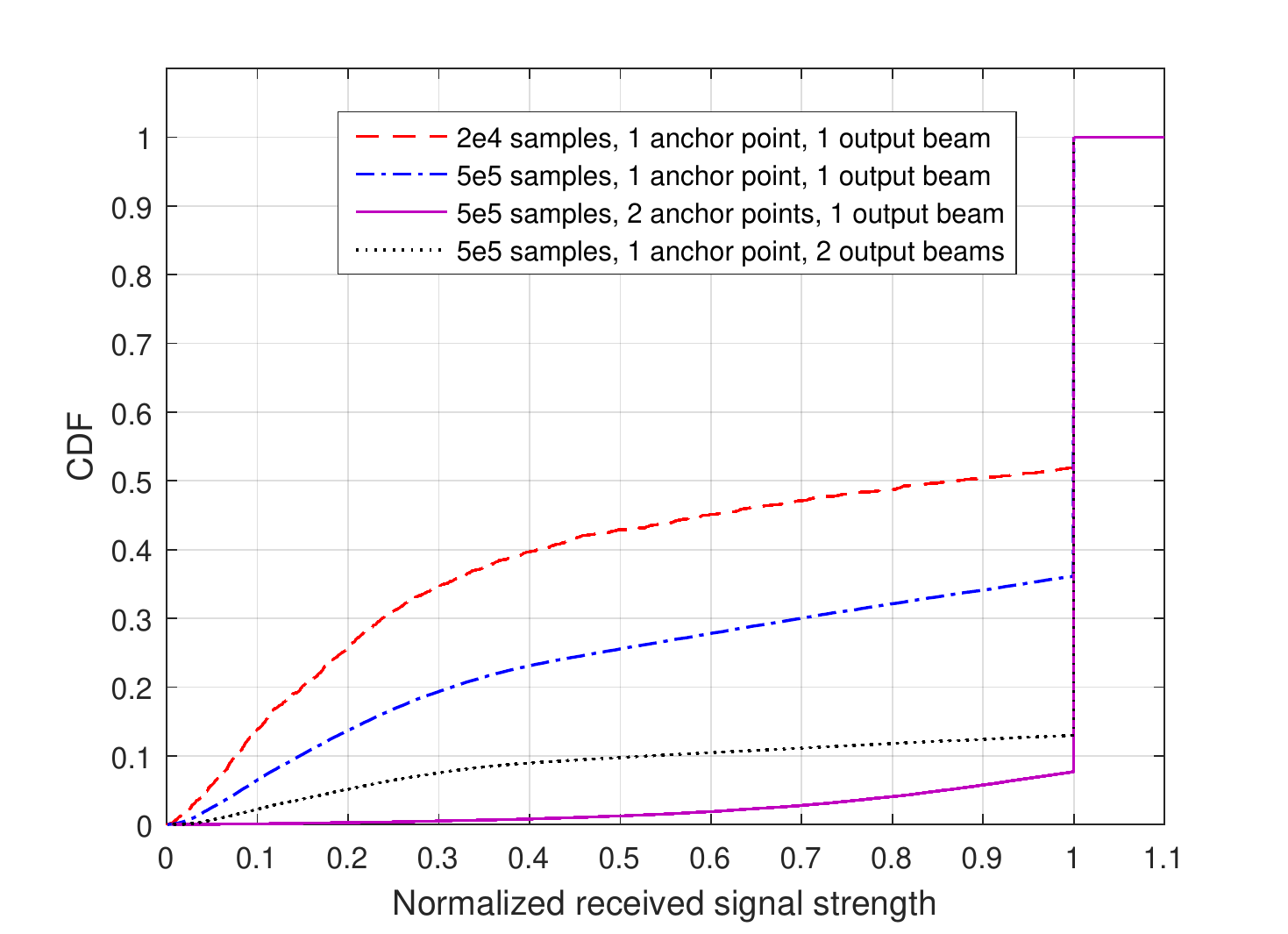}
\caption{CDF of normalized received signal strength compared with optimal beam pattern from the codebook.}
\label{sim_res1}
\end{figure}

Fig. \ref{sim_res1} shows the cumulative distribution function (CDF) of the normailzed received signal strength under different settings. With $20000$ samples, the accuracy is $48.1\%$  and the average normalized performance is $0.63$. When we increase the training sample size to $500000$, the prediction accuracy and average normalized performance are increased to $63.9\%$ and $0.77$, respectively. Moreover, we exploit the enhancements introduced in section \ref{section4}. It can be observed that including joint learning of multiple CBSs can indeed improve system performance. Simulation results also show that using more candidate beam patterns from network output also leads to a higher accuracy. In the case where all the aforementioned improvements methods are adopted, we can reach an accuracy of $99.74\%$, which is good enough for successful beamforming.

\begin{figure}[!t]
\centering
\includegraphics[width=3.5in]{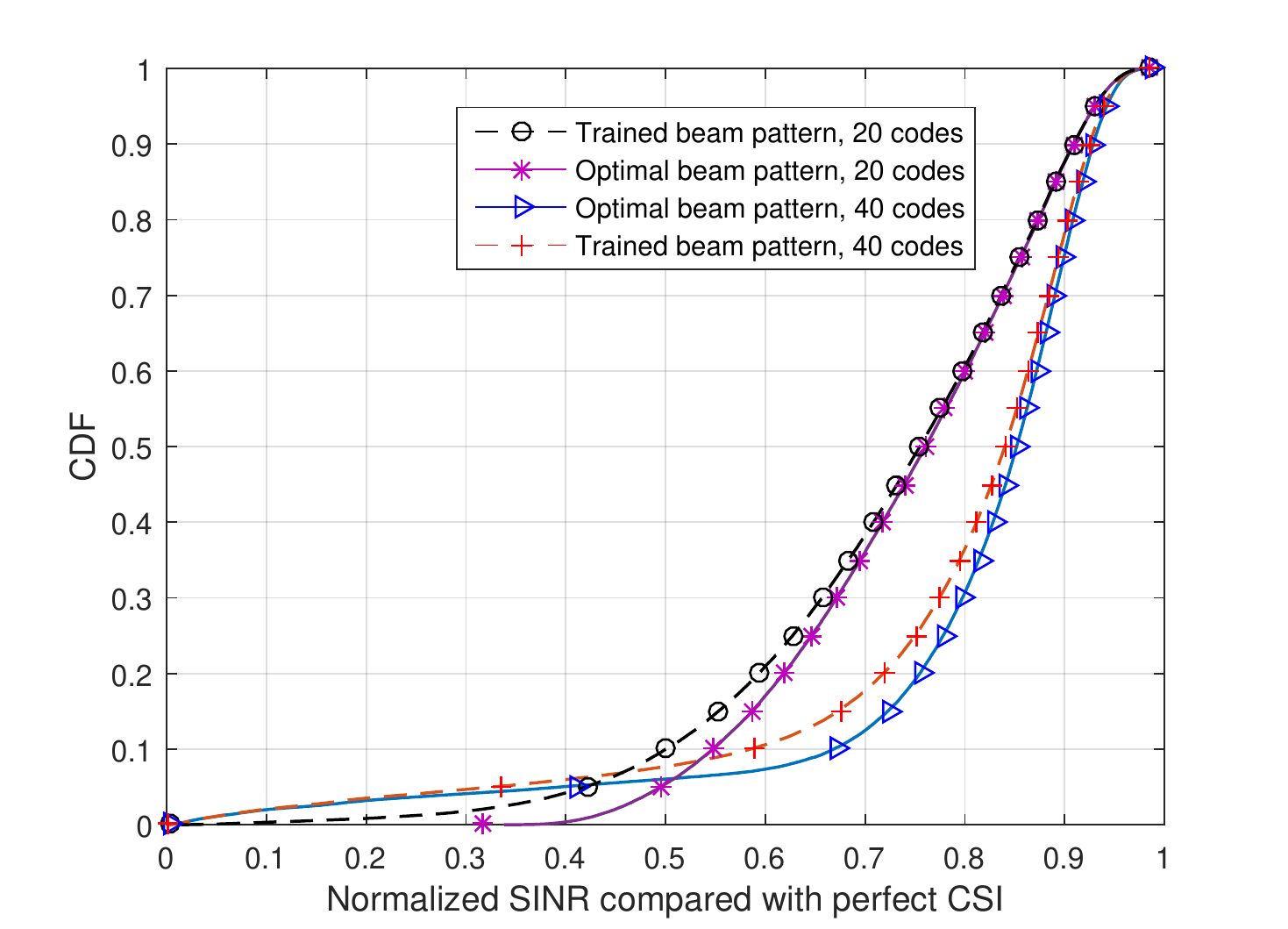}
\caption{CDF of normalized SINR compared with perfect CSI. We use the network trained by $500000$ samples and $2$ CBSs with two different codebook sizes. }
\label{sim_sinr}
\end{figure}

Fig. \ref{sim_sinr} gives a simulation of multiple-UE performance. To simplify the scenario, we assume that there are two UEs in the same TBS. We can infer the CSI from the trained beamforming vector and use minimum mean square error (MMSE) precoding to reduce inter-UE interference. It can be observed that the finite codebook quantization loss is the major loss.

\subsection{Comparisons with Location-Based Schemes}
\begin{figure}[!t]
\centering
\includegraphics[width=3.5in]{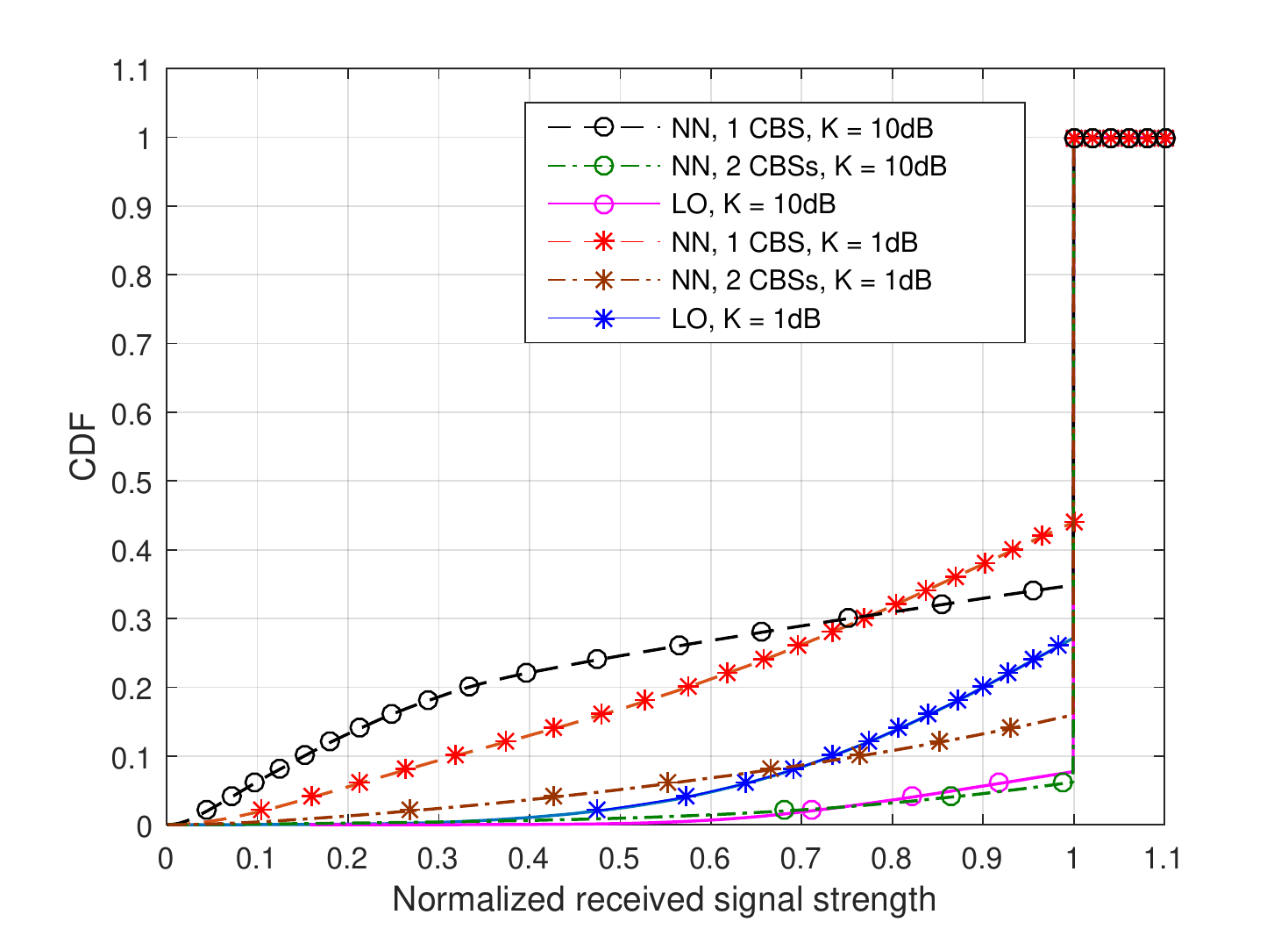}
\caption{Comparisons between location-based scheme (LO) and our beam inference scheme (NN) with different Rician factors and CBSs. Performance of location-based scheme is obtained with the knowledge of exact location information.}
\label{sim_com}
\end{figure}

We compare our work with location-based beamforming schemes. e.g., in \cite{kela16}. Existing location-based beamforming schemes estimate the locations of UEs and subsequently use directional vectors for beamforming. These schemes are highly dependent on the positioning accuracy and physical environments. Specifically, location-based beamforming is not capable to cope well with non-line-of-sight scenarios. With LoS scenario, location-based schemes still have beamforming error due to multi-path effects. Fig. \ref{sim_com} shows a comparison between the proposed scheme and a location-based scheme. Results of the location-based scheme are obtained under the assumption that user location is known exactly, which can be considered as a performance bound for location-based scheme. It can be observed that the performance of the proposed beam inference scheme with joint learning of two CBSs is almost as good as location-based algorithm. It is also worthwhile to stress that in reality, the locations of UEs may be difficult to obtain due to large positioning overhead and privacy issues.

\subsection{Influences of Other Factors}
\begin{figure}[!t]
\centering
\includegraphics[width=3.5in]{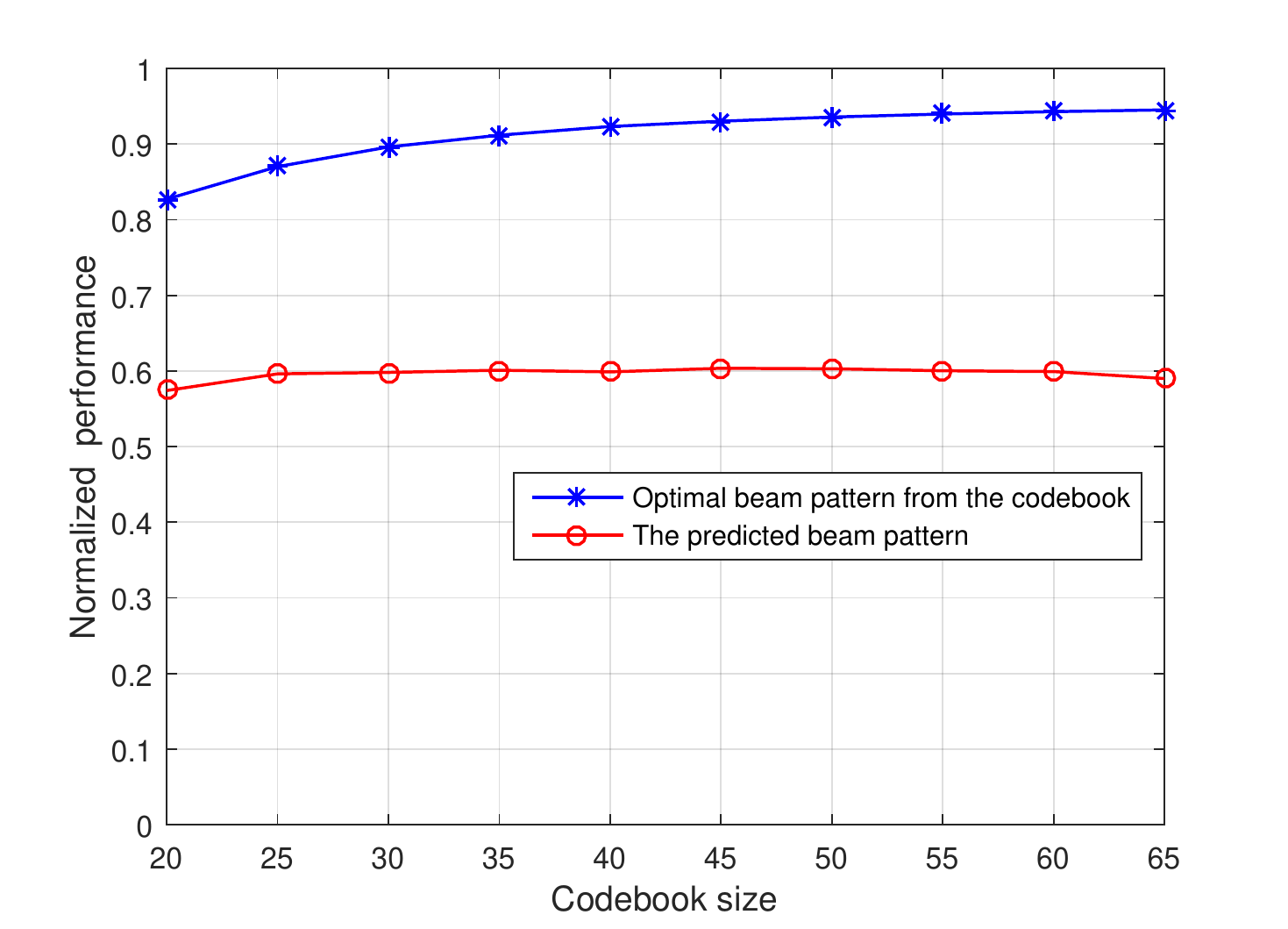}
\caption{Normalized performance v.s. codebook size. The number of training samples is $100000$.}
\label{sim_code}
\end{figure}

\begin{figure}[!t]
\centering
\includegraphics[width=3.5in]{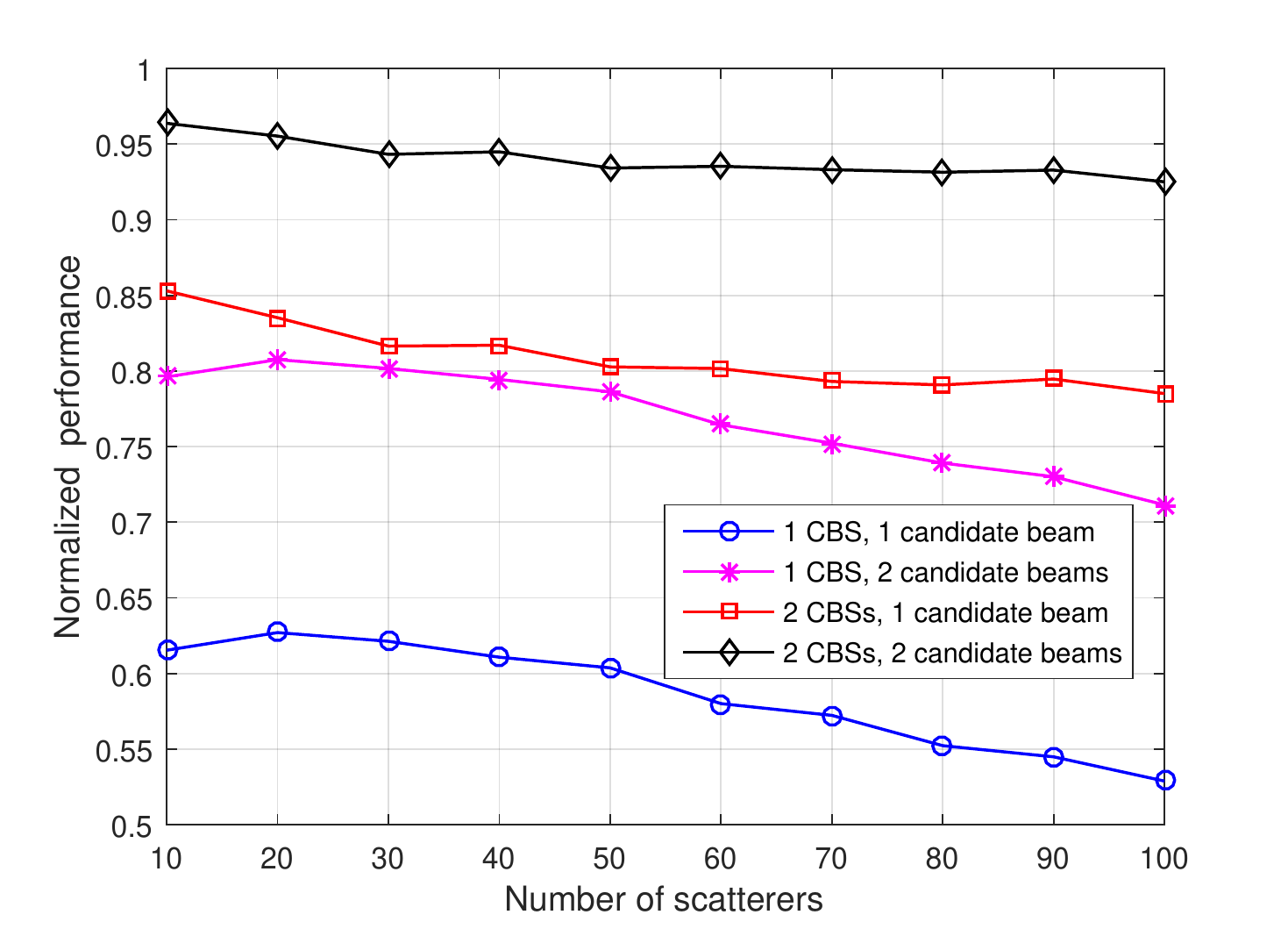}
\caption{Normalized performance v.s. number of scatterers. Simulation results are obtained from $20000$ samples with different number of CBSs and candidate beam patterns.}
\label{sim_sca}
\end{figure}

Other factors such as the codebook size and the number of scatterers also have an influence on system performance. Results in Fig. \ref{sim_code} show that with increasing codebook size, quantization loss is getting smaller but prediction error is growing larger. Therefore, there is a tradeoff between learning performance and codebook quantization loss. Fig. \ref{sim_sca} shows the relationship between average performance and the number of scatterers. It can be seen that although increasing number of scatterers leads to a performance loss, this kind of loss is not severe.

\section{Conclusion}
\label{section6}
In this paper, we propose a learning-based remote beamforming inference scheme in ultra-dense HCN. By calculating the mutual information between the CSI at CBS and the optimal beam pattern at TBS based on a data-driven approach, it is found that there exists a non-linear relationship between these two variables. Thereby, a supervised learning based channel inference method is proposed, in which the neural networks are leveraged to make the inference. Simulation results based on stochastic ray-tracing channel models are presented that a prediction accuracy of $63.9\%$ can be achieved by the trained network. It is also discovered that by joint learning of multiple CBSs and using more candidate beamforming patterns, the prediction can achieve an accuracy of $99.74\%$. More importantly, it can be concluded that based on our proposed framework, pilot resources and training overhead can be significantly reduced, or even eliminated for TBSs given the CSI at the CBS at the cost of building and training a neural  network in an offline fashion.

There are many promising directions based on this framework, such as scalable training methods for massive antenna arrays, improved performance with multiple users, qualitative analysis of system overhead, performance evaluation using field channel measurements and moving towards practical application.

\section*{Acknowledgment}
This work is sponsored in part by the Nature Science Foundation of China (No. 61461136004, No. 91638204, No. 61571265), and Intel Collaborative Research Institute for Mobile Networking and Computing.

\IEEEtriggeratref{7}
\bibliographystyle{ieeetr}
\bibliography{beam}

\end{document}